\documentclass[12pt]{article}
\usepackage{graphicx}

\newcommand\pubnumber{SNSN-323-63}
\newcommand\pubdate{\today}

\def\institute{Institute for Experimental Particle Physics\\
Karlsruhe Institute of Technology, 76131 Karlsruhe, GERMANY}
\def\support{\footnote{Supported by the Doctoral School “Karlsruhe School of Elementary and Astroparticle Physics: Science and Technology”}}

\def\Title#1{\begin{center} {\Large #1 } \end{center}}
\def\Author#1{\begin{center}{ \sc #1} \end{center}}
\def\Address#1{\begin{center}{ \it #1} \end{center}}

\newcommand\pubblock{\rightline{\begin{tabular}{l} \pubnumber\\
         \pubdate  \end{tabular}}}
\newenvironment{Abstract}{\begin{quotation}  }{\end{quotation}}
\newenvironment{Presented}{\begin{quotation} \begin{center} 
             PRESENTED AT\end{center}\bigskip 
      \begin{center}\begin{large}}{\end{large}\end{center} \end{quotation}}

\def\beq{\begin{equation}}
\def\eeq#1{\label{#1}\end{equation}}
\def\eeqn{\end{equation}}

\def\beqa{\begin{eqnarray}}
\def\eeqa#1{\label{#1}\end{eqnarray}}
\def\eeqan{\end{eqnarray}}

\let\bar=\overbar

\def\Dslash{\not{\hbox{\kern-4pt $D$}}}
\def\dslash{\not{\hbox{\kern-2pt $\del$}}}

\def\msb{{\bar{\ssstyle M \kern -1pt S}}}

\begin{document}
\begin{titlepage}
\pubblock

\vfill
\Title{Measurements of associated $\textrm{t}\overline{\textrm{t}}\textrm{X}$}
\vfill
\Author{Jan~van~der~Linden\support\\on behalf of the ATLAS and CMS Collaborations}
\Address{\institute}
\vfill
\begin{Abstract}
	Inclusive and differential cross section measurements of $\textrm{t}\overline{\textrm{t}}$-associated processes are performed at the ATLAS~\cite{ATLAS:2008xda} and CMS~\cite{CMS:2008xjf} Collaborations with high precision. In this contribution, a selection of recent measurements of inclusive and differential measurements of the $\textrm{t}\overline{\textrm{t}}\gamma$, $\textrm{t}\overline{\textrm{t}}\textrm{Z}$ and $\textrm{t}\overline{\textrm{t}}\textrm{W}$ processes are presented. Searches for $\textrm{t}\overline{\textrm{t}}\textrm{t}\overline{\textrm{t}}$ production are presented for both collaborations, providing evidence for the process.
\end{Abstract}
\vfill
\begin{Presented}
$15^\mathrm{th}$ International Workshop on Top Quark Physics\\
Durham, UK, 4--9 September, 2022
\end{Presented}
\vfill
\end{titlepage}
\def\thefootnote{\fnsymbol{footnote}}
\setcounter{footnote}{0}

\section{Introduction}

With the end of Run 2 of the LHC, sufficiently large datasets have been collected to enable high-precision inclusive and differential cross sections measurements of $\textrm{t}\overline{\textrm{t}}$-associated processes.
High purities of the $\textrm{t}\overline{\textrm{t}}\gamma$, $\textrm{t}\overline{\textrm{t}}\textrm{Z}$ and $\textrm{t}\overline{\textrm{t}}\textrm{W}$ states can be achieved in lepton-dominated final states. 
In the following, leptons refer to electrons or muons unless stated otherwise.
The large amount of data collected between 2015 and 2018 enables a measurement of these processes beyond the $10\%$ level of precision.\\
Searches for four top ($\textrm{t}\overline{\textrm{t}}\textrm{t}\overline{\textrm{t}}$) production are performed by both the ATLAS and CMS Collaborations. This process suffers from small signal cross sections and difficult backgrounds, but nevertheless evidence in measurements by both experiments has been found.\\
In this contribution only a selection of recent $\textrm{t}\overline{\textrm{t}}\gamma$, $\textrm{t}\overline{\textrm{t}}\textrm{Z}$, $\textrm{t}\overline{\textrm{t}}\textrm{W}$ and $\textrm{t}\overline{\textrm{t}}\textrm{t}\overline{\textrm{t}}$ measurements are presented.
Table~\ref{tab:summary} summarizes the inclusive cross section measurements discussed here.

\begin{table}[b!]
\begin{center}
	\caption{Summary of inclusive cross section measurements discussed in this contribution, rounded to fb. The values are compared to predictions, either from dedicated calculations where referenced, or from Monte-Carlo event generator predictions where no reference is given.}
	\label{tab:summary}
\begin{tabular}{lcclcl}  
\hline\hline\\[-3mm]
\textbf{Process} & \textbf{Experiment} & \multicolumn{2}{c}{\textbf{Measurement}}  & \multicolumn{2}{c}{\textbf{Prediction}}  \\[1mm] 
\hline\\[-3mm]
$\textrm{t}\overline{\textrm{t}}\gamma\ (2\ell)$ & CMS & $175\pm 7\,\textrm{fb}$ & \cite{CMS:2022lmh} & $155\pm27\,\textrm{fb}$ & \\[1mm]

 $\textrm{t}\overline{\textrm{t}}\textrm{Z}\ (3\ell/4\ell)$ & ATLAS & $990\pm90\,\textrm{fb}$ & \cite{ATLAS:2021fzm} & $859\pm78\,\textrm{fb}$ & \cite{Kulesza:2018tqz} \\[1mm]
 
 $\textrm{t}\overline{\textrm{t}}\textrm{W}\ (2\ell/3\ell)$ & CMS & $870\pm60\,\textrm{fb}$ & \cite{CMS:2022tkv} & $722\pm74\,\textrm{fb}$& \cite{Frederix:2021agh} \\[1mm]
 
 $\textrm{t}\overline{\textrm{t}}\textrm{t}\overline{\textrm{t}}\ (\geq1\ell)$ & ATLAS & $24\pm7\,\textrm{fb}$ & \cite{ATLAS:2020hpj,ATLAS:2021kqb} & $12\pm2\,\textrm{fb}$  & \cite{Frederix:2017wme}\\[1mm]
 $\textrm{t}\overline{\textrm{t}}\textrm{t}\overline{\textrm{t}}$ & CMS & $17\pm5\,\textrm{fb}$ & \cite{CMS:2022uga,CMS:2019rvj,CMS:2019jsc} & $12\pm2\,\textrm{fb}$  & \cite{Frederix:2017wme}\\[1mm]
 \hline\hline
\end{tabular}
\end{center}
\end{table}

\section{$\textrm{t}\overline{\textrm{t}}\gamma$ measurements}
\begin{figure}[t]
	\centering
	\includegraphics[height=2.3in]{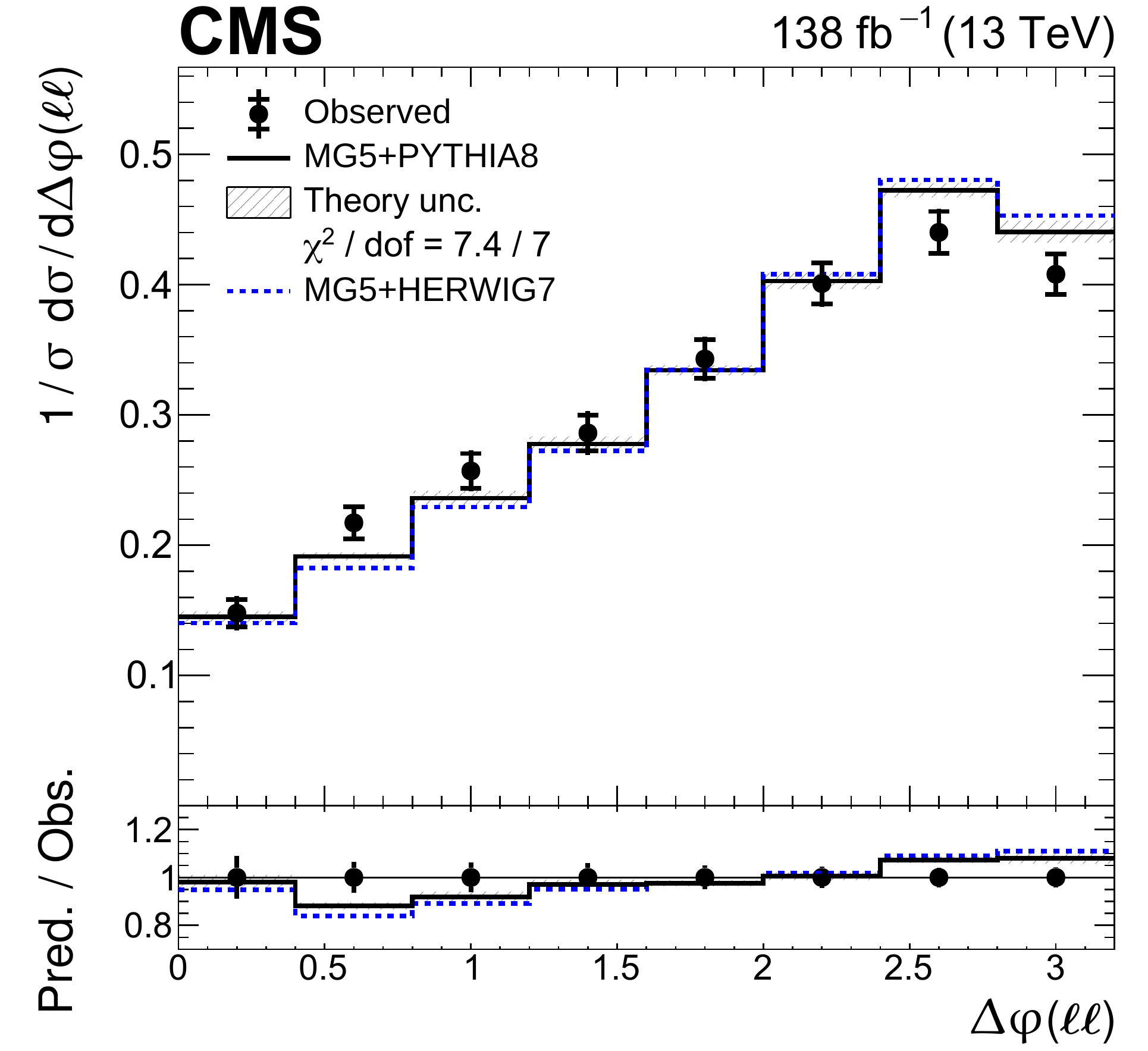}
	\includegraphics[height=2.3in]{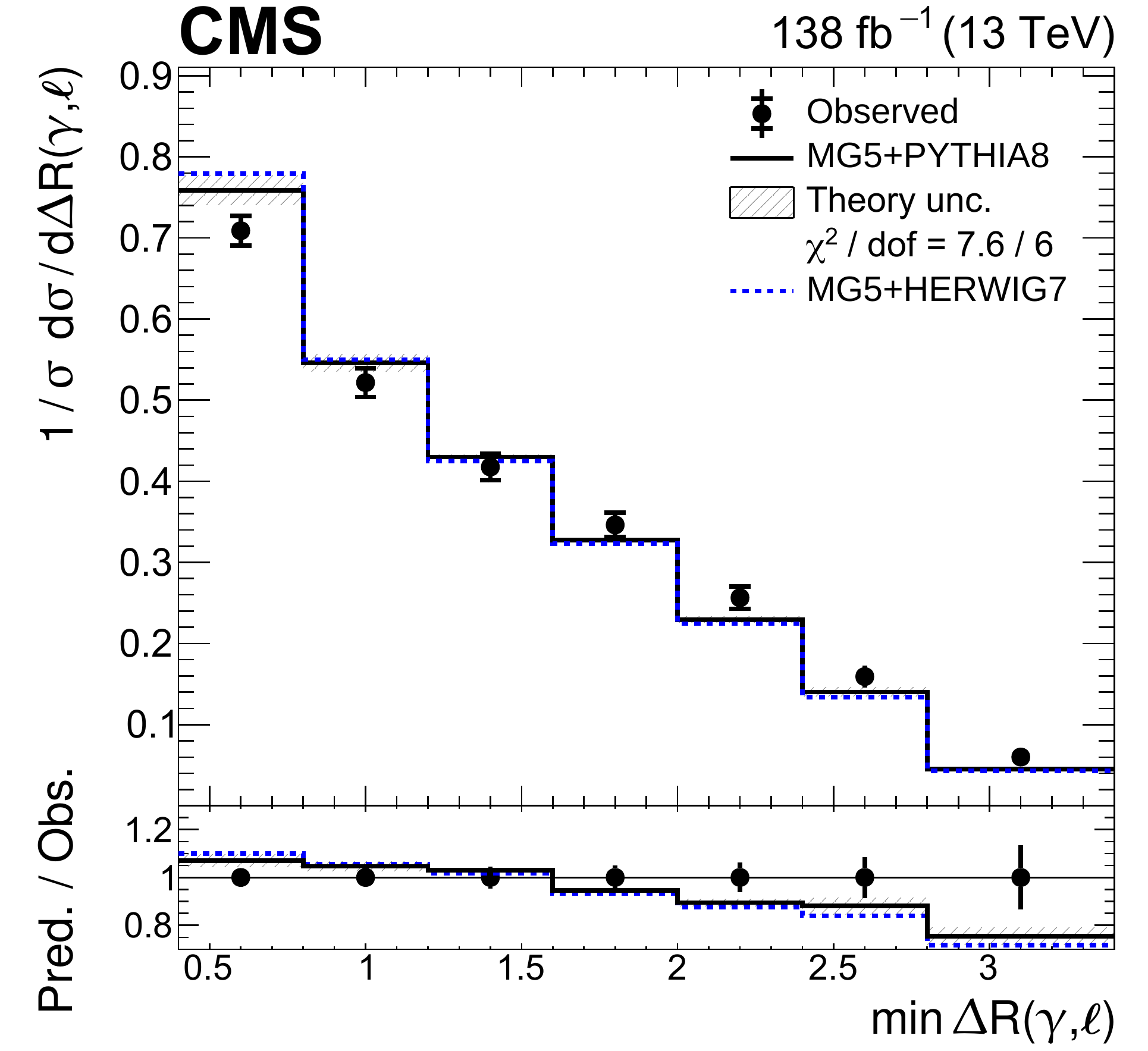}
	\caption{Normalized differential $\textrm{t}\overline{\textrm{t}}\gamma$ cross section measurements as a function of the azimuthal angle between the two leptons (left) and the minimum $\Delta R$ between $\gamma$ and lepton (right)~\cite{CMS:2022lmh}.}
	\label{fig:ttgamma}
\end{figure}

The CMS Collaboration measures the $\textrm{t}\overline{\textrm{t}}\gamma$ cross section in events with two leptons of opposite charge using $138\,\textrm{fb}^{-1}$ of data~\cite{CMS:2022lmh}. A fiducial region is defined at particle level to define the $\textrm{t}\overline{\textrm{t}}\gamma$ process.
A template fit is performed to the $p_{\textrm{\scriptsize{T}}}$ distribution of the reconstructed photon to extract the fiducial cross section with a precision of $4\%$. The result is around 10\% higher compared to Monte-Carlo event generator predictions.
Differential measurements are performed for a wide range of photon kinematics, lepton-related observables and lepton+$\gamma$ observables. The distributions are unfolded to particle level.
Figure~\ref{fig:ttgamma} displays two representative distributions.
Trends between the measurement and the simulation can be observed, hinting towards mismodelings of the $\gamma$ origin ($\gamma$ radiation from production or decay) in the simulation.\\
A measurement of the charge asymmetry in $\textrm{t}\overline{\textrm{t}}\gamma$ events in the single lepton channel is performed by the ATLAS Collaboration using $139\,\textrm{fb}^{-1}$ of data~\cite{ATLAS:2022gvp}.
The measurement probes the charge asymmetries of top quarks and antiquarks in $\textrm{t}\overline{\textrm{t}}\gamma$ production, therefore $\gamma$ radiation in $\textrm{t}\overline{\textrm{t}}$ decay is treated as background.
A deep neural network (DNN) is employed to identify $\textrm{t}\overline{\textrm{t}}\gamma$ production. The output of the DNN is used to define a signal- and background-enriched region. In each of the regions the sign of $\Delta y_{\textrm{t}}$, the difference in rapidity $y$ of the top quark and antiquark, is used in a template fit to data to determine the charge asymmetry $A_{\textrm{c}}$.
The top quark and antiquark are identified via kinematic reconstruction.
The measured value is $A_{\textrm{c}} = -0.006 \pm 0.024\textrm{(stat)} \pm 0.018 \textrm{(syst)}$.
This result is compatible with the expectation of $A_{\textrm{c}} = -0.014 \pm 0.001\textrm{(scale)}$ from simulation.

\section{$\textrm{t}\overline{\textrm{t}}\textrm{Z}$ measurements}
\begin{figure}[t]
	\centering
	\includegraphics[height=2.3in]{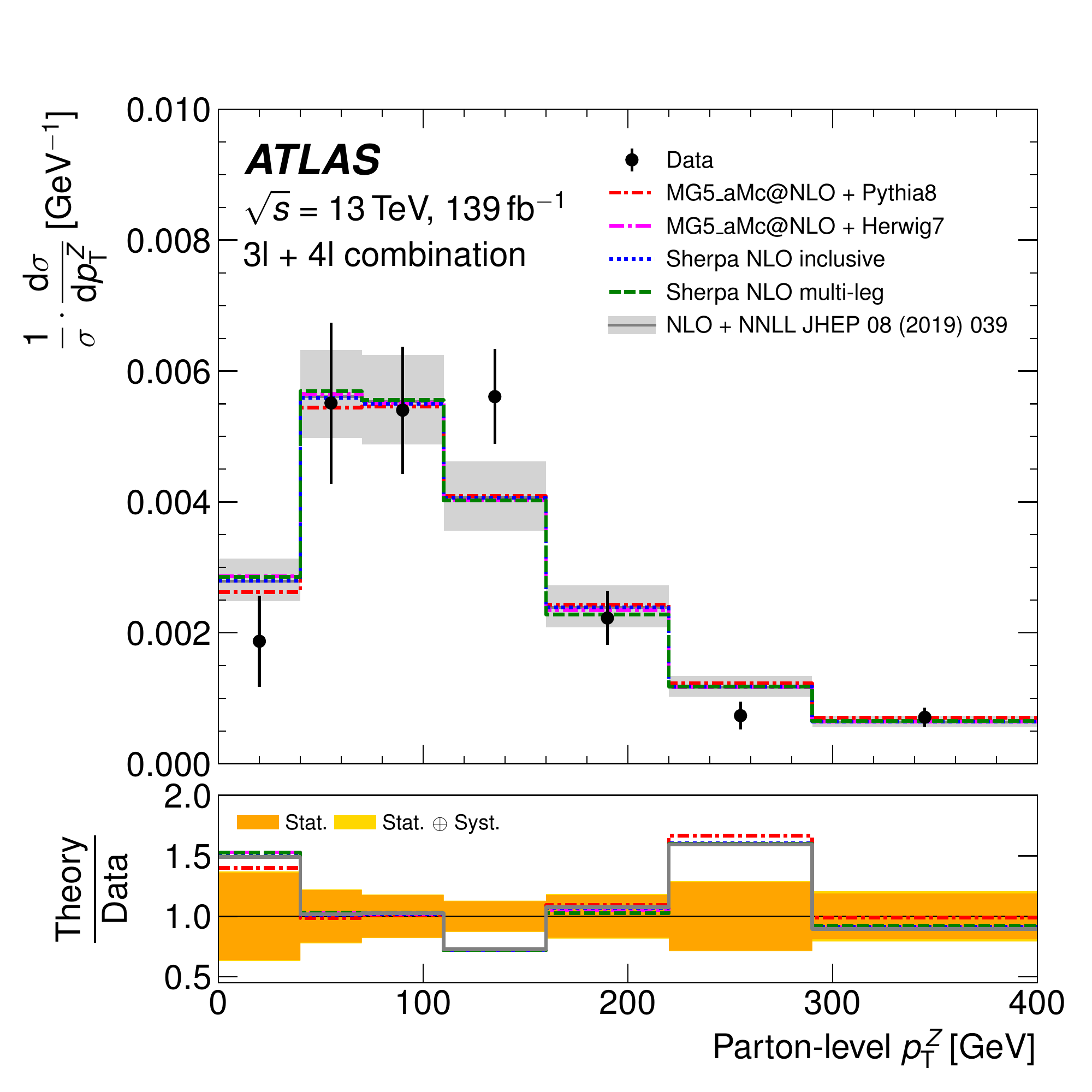}
	\includegraphics[height=2.0in]{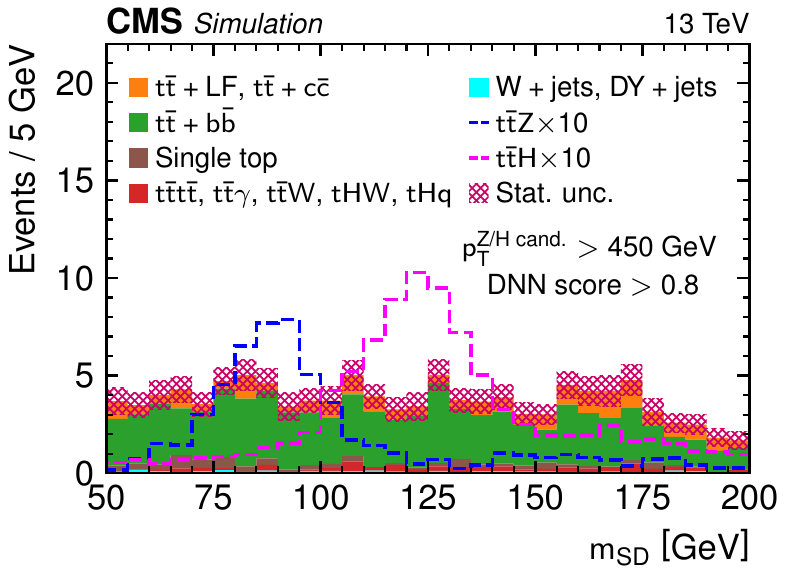}
	\caption{Normalized differential $\textrm{t}\overline{\textrm{t}}\textrm{Z}$ cross section measurement as a function of the parton-level $p_{\textrm{\textrm{\scriptsize{T}}}}$ of the Z boson (left)~\cite{ATLAS:2021fzm}. Soft-drop mass distribution of candidate Higgs or Z boson used in a fit to data (right) to extract the $\textrm{t}\overline{\textrm{t}}\textrm{Z}$ and $\textrm{t}\overline{\textrm{t}}\textrm{H}$ cross sections~\cite{CMS:2022hjj}.}
	\label{fig:ttz}
\end{figure}

A measurement of the $\textrm{t}\overline{\textrm{t}}\textrm{Z}$ cross section is performed by the ATLAS Collaboration in final states with three or four charged leptons using $139\,\textrm{fb}^{-1}$ of data~\cite{ATLAS:2021fzm}.
The cross section is extracted via a template fit to data in bins of lepton, jet and b jet multiplicities.
The cross section is measured with a precision of $10\%$.
Differential distributions are measured via unfolding to particle and parton level.
Observables are defined to probe the top-Z coupling, as well as the modeling of Z boson, top quark and additional radiation.
Figure~\ref{fig:ttz} (left) shows one representative observable.\\
The CMS Collaboration measures the $\textrm{t}\overline{\textrm{t}}\textrm{Z}$ and $\textrm{t}\overline{\textrm{t}}\textrm{H}$ cross sections in a boosted regime using $138\,\textrm{fb}^{-1}$ of data~\cite{CMS:2022hjj}.
The measurement targets the $\textrm{b}\overline{\textrm{b}}$ final state of the Higgs and Z boson in a single lepton channel.
Large-radius (AK8) jets are tagged as Higgs or Z boson resonances. Additionally, DNNs are used to separate the signal processes from backgrounds. 
A template fit to the AK8 jet soft-drop mass, the DNN score and the $p_\textrm{\scriptsize{T}}$ of the AK8 jet is performed to data to extract both cross sections.
Figure~\ref{fig:ttz} (right) shows the soft-drop mass in a representative $p_{\textrm{\scriptsize{T}}}$ and DNN range used in the fit.
The $\textrm{t}\overline{\textrm{t}}\textrm{Z}$ and $\textrm{t}\overline{\textrm{t}}\textrm{H}$ cross sections are measured with a low correlation of $-10\%$. The sensitivity is limited by data statistics.

\section{$\textrm{t}\overline{\textrm{t}}\textrm{W}$ measurements}
\begin{figure}[t]
	\centering
	\includegraphics[height=2.3in]{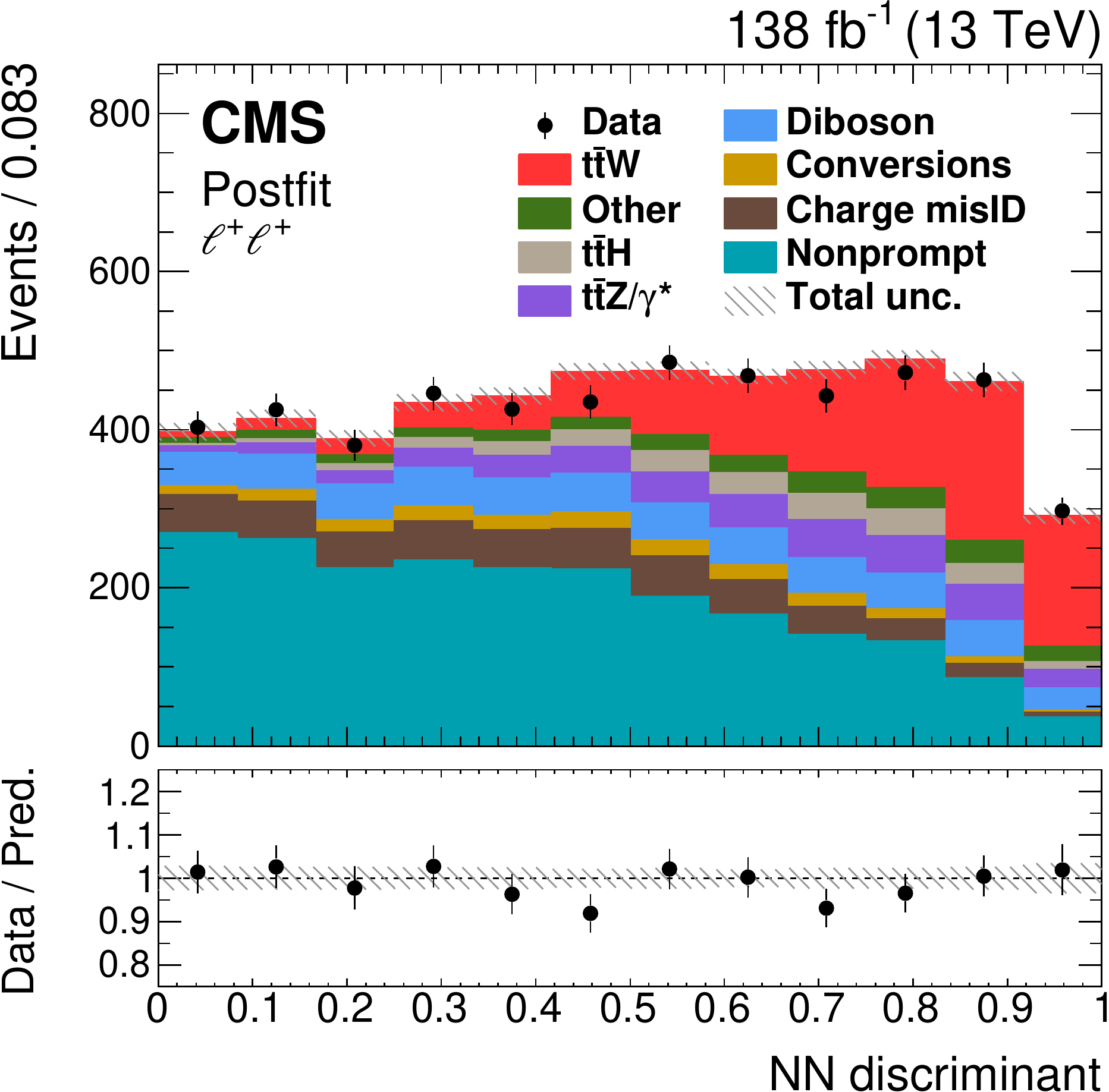}
	\includegraphics[height=2.3in]{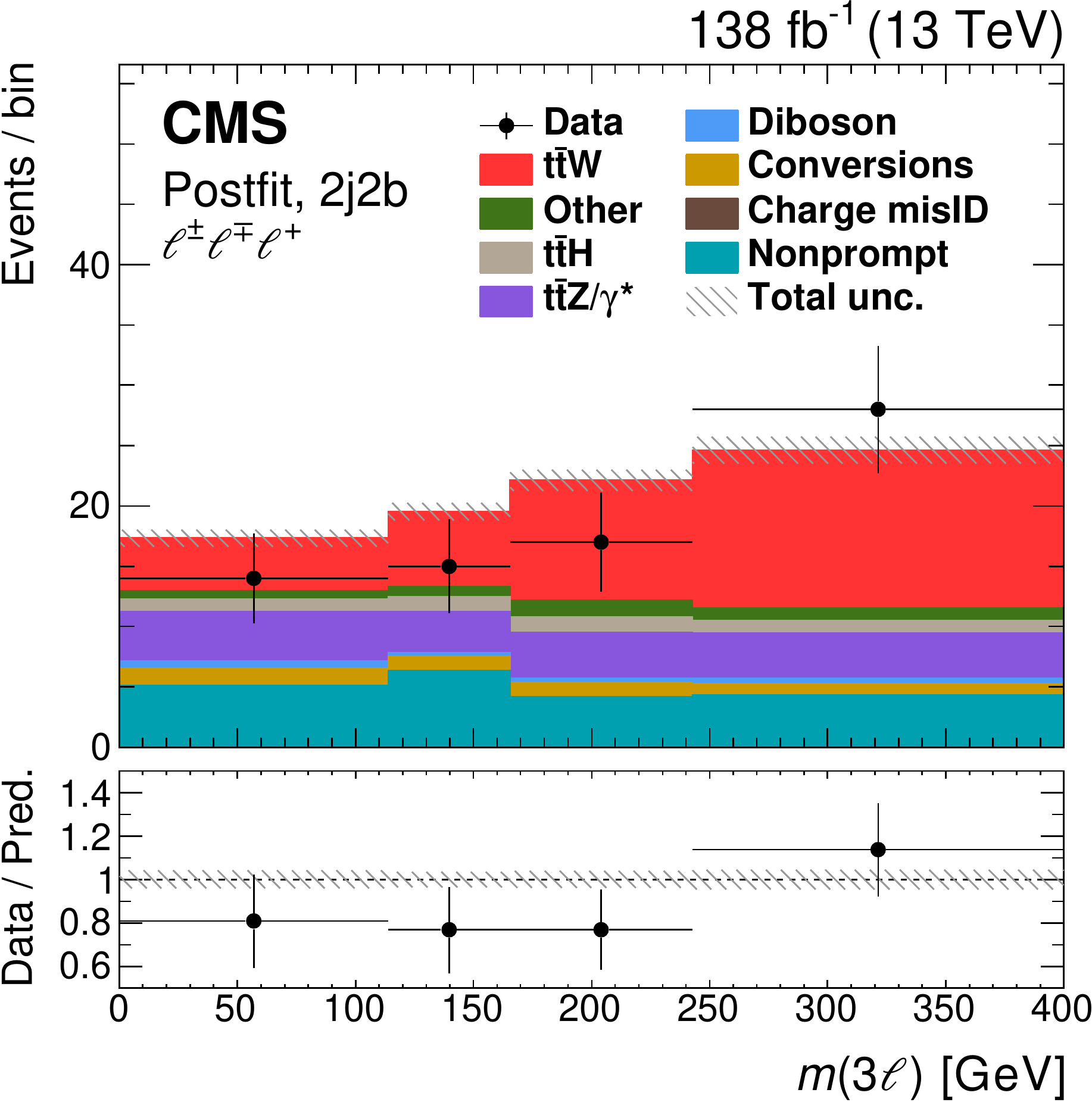}
	\caption{DNN score in dilepton channel (left) and invariant mass of three leptons (right) used in a fit to data to extract the $\textrm{t}\overline{\textrm{t}}\textrm{W}$ cross section~\cite{CMS:2022tkv}.}
	\label{fig:ttw}
\end{figure}

A measurement of the $\textrm{t}\overline{\textrm{t}}\textrm{W}$ cross section is performed by the CMS Collaboration in final states with a same-sign dilepton pair or three leptons using $138\,\textrm{fb}^{-1}$ of data~\cite{CMS:2022tkv}.
In the region with two leptons a DNN is employed to separate the signal process from backgrounds.
In the region with three leptons events are separated based on the jet and b jet multiplicity.
The $\textrm{t}\overline{\textrm{t}}\textrm{W}$ cross section is extracted in a template fit to data using the DNN score in the two-lepton region and the invariant mass of all leptons in the three-lepton region.
Figure~\ref{fig:ttw} shows two representative distributions used in the fit.
The $\textrm{t}\overline{\textrm{t}}\textrm{W}$ cross section is measured with a precision of $10\%$, showing a trend to higher values compared to NLO+FxFx calculations~\cite{Frederix:2021agh}.
Additionally, measurements for the separate $\textrm{t}\overline{\textrm{t}}\textrm{W}^{+}$ and $\textrm{t}\overline{\textrm{t}}\textrm{W}^{-}$ cross sections and their ratio are provided, all of them also showing values higher than the calculations.\\
A measurement of the charge asymmetry of top quarks and antiquarks is performed by the ATLAS Collaboration in $\textrm{t}\overline{\textrm{t}}\textrm{W}$ events with three leptons in the final state using $139\,\textrm{fb}^{-1}$ of data~\cite{ATLAS:2022sds}.
Events are separated in bins with positive or negative $\Delta \eta^\ell$, the pseudo rapidity difference of leptons from both top quark decays. 
A BDT is used to identify the two leptons from top quark and antiquark decays.
Unfolding of the distribution to particle level is performed to access the rapidity difference of top quark and antiquark.
The resulting value of the charge asymmetry is $A_{\textrm{c}} = -0.112 \pm 0.170\textrm{(stat)} \pm 0.055 \textrm{(syst)}$.
This is compatible with the expectation of  $A_{\textrm{c}} = -0.063 \pm 0.008\textrm{(scale+MC stat)}$ from simulation.

\section{$\textrm{t}\overline{\textrm{t}}\textrm{t}\overline{\textrm{t}}$ searches}

\begin{figure}[t]
	\centering
	\includegraphics[height=2.3in]{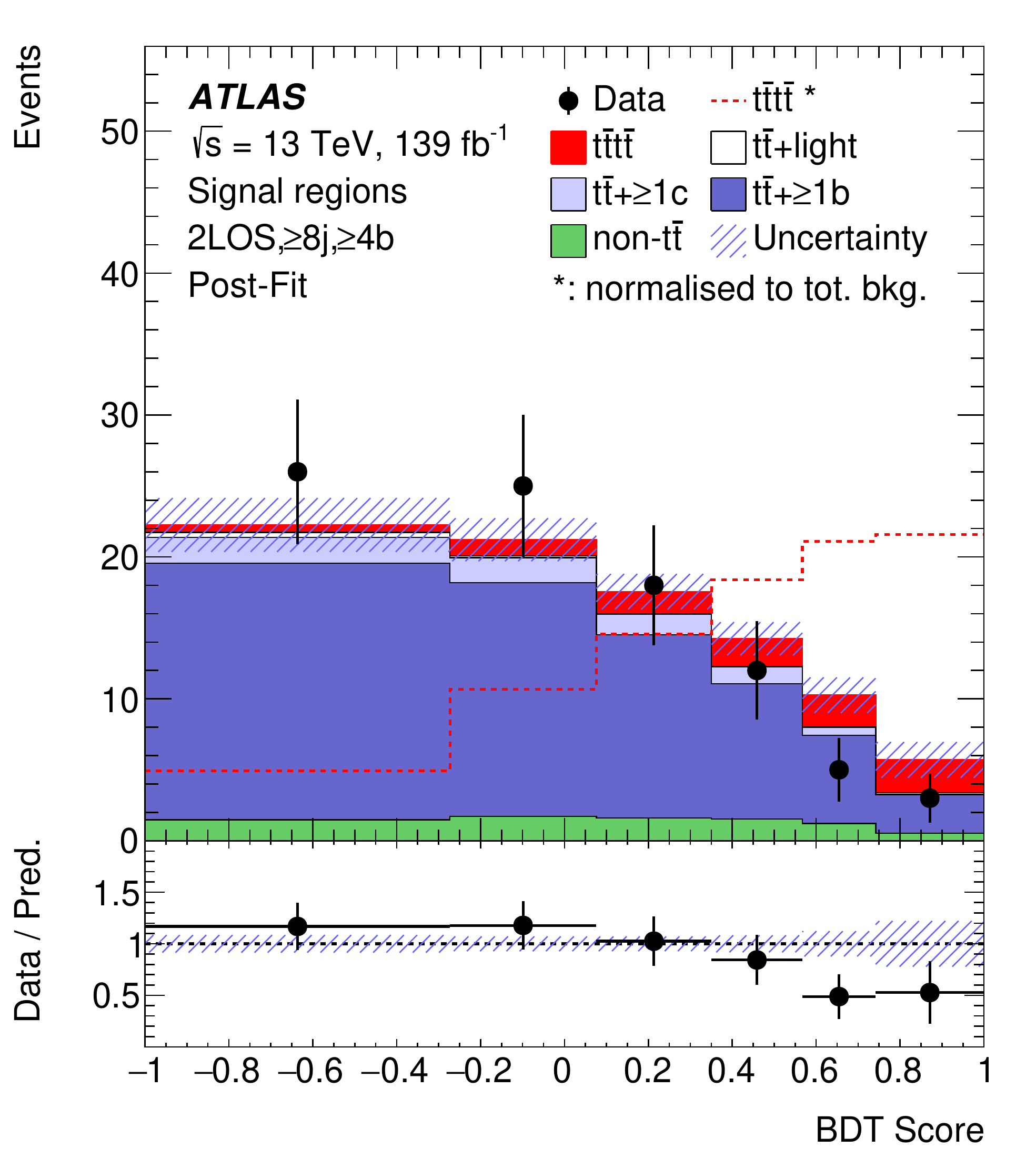}
	\includegraphics[height=2.3in]{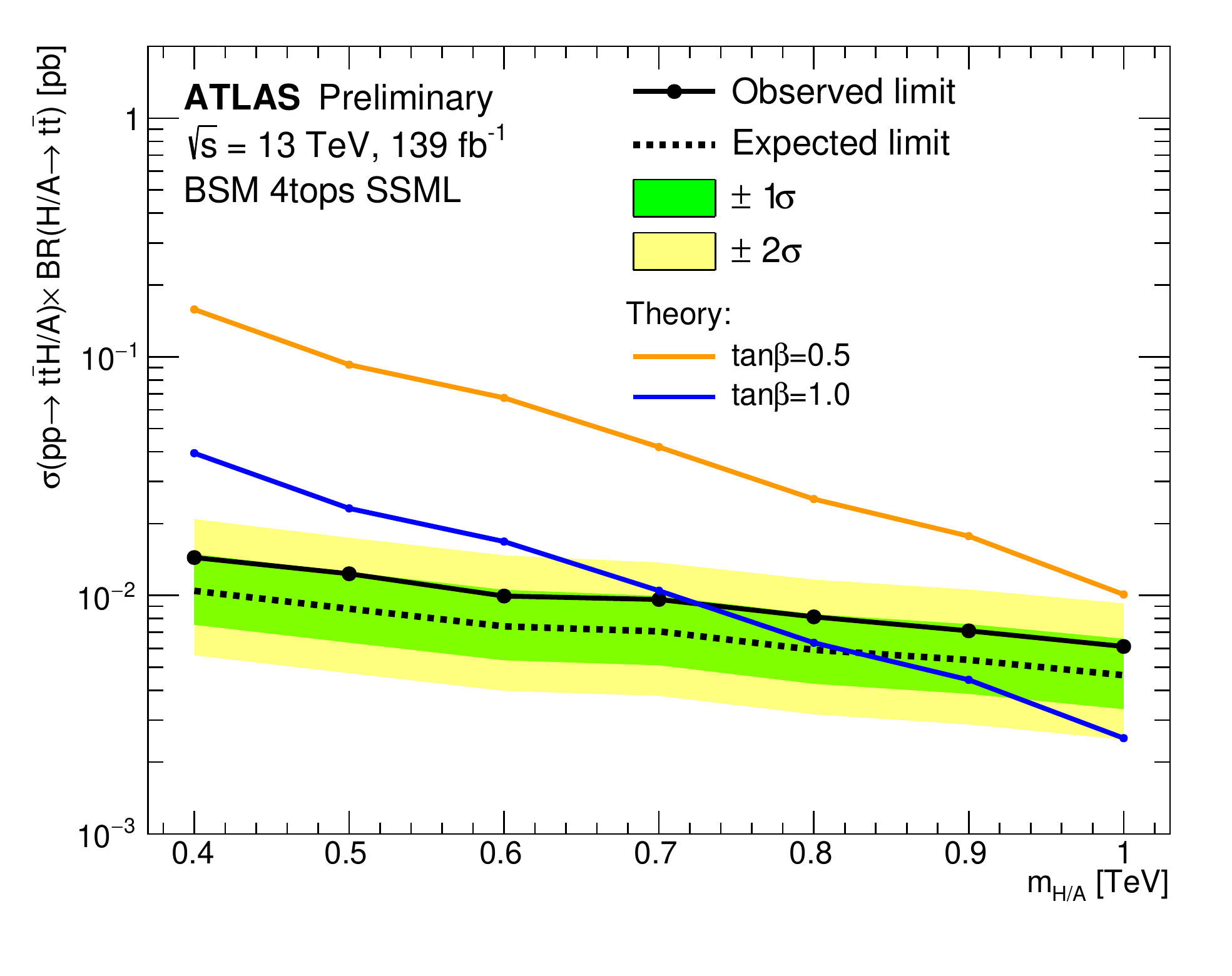}
	\caption{BDT score in the dilepton channel used in a fit to data to extract the $\textrm{t}\overline{\textrm{t}}\textrm{t}\overline{\textrm{t}}$ cross section (left)~\cite{ATLAS:2021kqb}. Exclusion limits of heavy $\textrm{H}/\textrm{A}$ in a mass range of $400\,\textrm{GeV}$ to $1000\,\textrm{GeV}$ (right)~\cite{ATLAS:2022ohr}.}
	\label{fig:ttttatlas}
\end{figure}

A search for $\textrm{t}\overline{\textrm{t}}\textrm{t}\overline{\textrm{t}}$ production is performed by the ATLAS Collaboration using $139\,\textrm{fb}^{-1}$ of data~\cite{ATLAS:2020hpj}. The search is performed in final states with a same-sign dilepton pair or more than two leptons.
A BDT is employed to separate signal and background processes. A template fit is performed to the output values of the BDT.\\
A separate search for $\textrm{t}\overline{\textrm{t}}\textrm{t}\overline{\textrm{t}}$ production is performed by the ATLAS experiment in final states with an opposite-sign dilepton pair or a single lepton~\cite{ATLAS:2021kqb}.
Events are separated into jet and b jet regions. The $\textrm{t}\overline{\textrm{t}}\textrm{t}\overline{\textrm{t}}$ cross section is extracted in a template fit to BDT and $H_\textrm{\textrm{\scriptsize{T}}}$ distributions, one representative shown in Figure~\ref{fig:ttttatlas} (left).\\
Both measurements are combined to yield a significance of $4.7$ standard deviations for the $\textrm{t}\overline{\textrm{t}}\textrm{t}\overline{\textrm{t}}$ production process.
The cross section measured in data is twice as high as in NLO QCD+EW calculations~\cite{Frederix:2017wme}.\\
A search for $\textrm{t}\overline{\textrm{t}}\textrm{H}/\textrm{A} \rightarrow \textrm{t}\overline{\textrm{t}}\textrm{t}\overline{\textrm{t}}$ is performed by the ATLAS Collaboration using $139\,\textrm{fb}^{-1}$ of data~\cite{ATLAS:2022ohr}. 
Here, $\textrm{H}/\textrm{A}$ describes a new heavy scalar boson which is analyzed in a decay to $\textrm{t}\overline{\textrm{t}}$.
The analysis strategy follows the strategy employed in the same-sign dilepton $\textrm{t}\overline{\textrm{t}}\textrm{t}\overline{\textrm{t}}$ search~\cite{ATLAS:2021kqb}. 
The same BDT is used for identification of $\textrm{t}\overline{\textrm{t}}\textrm{t}\overline{\textrm{t}}$ signatures, followed by an additional mass-parameterized BDT for identification of BSM signatures of variable masses.
Limits are set on $\textrm{H}/\textrm{A}$ masses in the range of $400\,\textrm{GeV}$ to $1000\,\textrm{GeV}$, as shown in Figure~\ref{fig:ttttatlas} (right).

A search for $\textrm{t}\overline{\textrm{t}}\textrm{t}\overline{\textrm{t}}$ production is performed by the CMS Collaboration using up to $138\,\textrm{fb}^{-1}$ of data~\cite{CMS:2022uga}.
The measurement is performed in final states with either an opposite-sign dilepton pair, a single lepton, and, for the first time, without leptons.
In the dilepton channel the $H_{\textrm{\scriptsize{T}}}$ distribution, separated by lepton flavour, jet and b jet multiplicity is used in a template fit to data.
In the single lepton channel a BDT is used for signal and background separation. The output of the BDT, separated by lepton flavour, jet and b jet multiplicity and number of boosted top quark candidates is used in a template fit to data.
In the channel without leptons a BDT is employed to separate signal and multijet signatures. 
Categories of resolved and boosted top candidate multiplicities, as well as $H_\textrm{\textrm{\scriptsize{T}}}$ are defined and consequently used in a template fit to data. Data-driven background estimations for contributions of multijet and $\textrm{t}\overline{\textrm{t}}{+}\textrm{jets}$ events are employed in this channel.\\
The template fit to data, combined with two previous measurements~\cite{CMS:2019rvj, CMS:2019jsc}, yields a cross section for $\textrm{t}\overline{\textrm{t}}\textrm{t}\overline{\textrm{t}}$ production $40\%$ higher than predicted by NLO QCD+EW calculations~\cite{Frederix:2017wme}. 
This combined measurement achieves a significance of $4.0$ standard deviations.

\end{document}